\documentclass[aps,prd,showpacs,twocolumn,superscriptaddress]{revtex4}
\usepackage{graphicx,amsmath,amsfonts}

\RequirePackage{xspace}

\newcommand{\BABARPubYear}     {05}
\newcommand{\BABARPubNumber}  {021}
\newcommand{\SLACPubNumber} {11287}

\input pubboard/babarsym.tex

\def\BaBar{\mbox{\slshape B\kern-0.1em{\smaller A}\kern-0.1em
    B\kern-0.1em{\smaller A\kern-0.2em R}}\xspace}
\def\babar{\BaBar}

\def\etal{{\it et al.}}

\newcommand{\bfg}{\begin{figure}}
\newcommand{\efg}{\end{figure}}
\newcommand{\infg}{\includegraphics}

\newcommand{\btbl}{\begin{table}}
\newcommand{\etbl}{\end{table}}
\newcommand{\btbu}{\begin{tabular}} 
\newcommand{\etbu}{\end{tabular}}

\newcommand{\abs}[1]{\lvert#1\rvert}

\newcommand{\mcol}[2]{\multicolumn{#1}{c|}{#2}}

\def\jsi{\jpsi}
\def\psip{\psitwos}
\def\etac{\eta_c(1S)\xspace}
\def\etacp{\eta_c(2S)\xspace}
\def\ee{e^+e^-}
\def\mm{\mu^+\mu^-}
\def\ll{\ell^+\ell^-}
\def\ccbar{c\bar{c}}
\def\ccThreeChg{\ccbar\to\,>2\,\mathrm{charged}}
\def\BR{{\cal{B}}}
\def\LUMI{{\cal{L}}}
\def\eff{\varepsilon}
\def\mrec{M_\textrm{rec}}
\def\cosHel{\cos\theta_l}
\def\jee{\jpsi\to\ee}
\def\jmm{\jpsi\to\mm}

\def\emcTotLess{E_\textrm{QED}-E_\textrm{beams}}
\def\Eqed{E_\textrm{QED}}

\graphicspath{{./eps/}}

\begin{document}

\preprint{\babar-PUB-\BABARPubYear/\BABARPubNumber}
\preprint{SLAC-PUB-\SLACPubNumber}

\begin{flushleft}
  \babar-PUB-\BABARPubYear/\BABARPubNumber\\
  SLAC-PUB-\SLACPubNumber\\
\end{flushleft}

\title{ \large \bf \boldmath
  Measurement of Double Charmonium Production in $\ee$ Annihilations 
  at $\sqrt{s}=10.6$~GeV
}

%
\author{B.~Aubert}
\author{R.~Barate}
\author{D.~Boutigny}
\author{F.~Couderc}
\author{Y.~Karyotakis}
\author{J.~P.~Lees}
\author{V.~Poireau}
\author{V.~Tisserand}
\author{A.~Zghiche}
\affiliation{Laboratoire de Physique des Particules, F-74941 Annecy-le-Vieux, France }
\author{E.~Grauges}
\affiliation{IFAE, Universitat Autonoma de Barcelona, E-08193 Bellaterra, Barcelona, Spain }
% ...
\collaboration{The \babar\ Collaboration}
\noaffiliation

\date{\today}

\begin{abstract}

  We study $\ee\to J/\psi\,\ccbar$ by measuring the 
  invariant mass distribution recoiling 
  against fully reconstructed $\jpsi$ decays, using 
  124\invfb of data collected with a center-of-mass energy of 10.6\gev
  with the $\BaBar$ detector. 
  We observe signals for $\etac$, $\chiczero$, and $\etacp$ 
  in the recoil mass distribution, thus confirming previous measurements. 
  We measure $\sigma(\ee\to\jpsi+\ccbar)\,\BR(\ccThreeChg)$ to be
  $17.6\pm2.8\stat^{+1.5}_{-2.1}\syst$~fb, 
  $10.3\pm2.5\stat^{+1.4}_{-1.8}\syst$~fb, and 
  $16.4\pm3.7\stat^{+2.4}_{-3.0}\syst$~fb 
  with $\ccbar=\etac$, $\chiczero$, and $\etacp$, respectively. 

\end{abstract}


\maketitle

Prompt $\jpsi$ and $\psip$ production in $\ee$ annihilations around
$\sqrt{s}=10.6$\gev has been observed by both 
the \BaBar~\cite{ct:BaBar-jsi} and Belle~\cite{ct:Belle-psi} experiments.
These interactions provide an opportunity to 
study both perturbative and non-perturbative effects in QCD 
and to search for new charmonium states~\cite
{ct:search-charmonium-1,ct:singleGamma-Chao}. 

Belle~\cite{ct:Belle-double_ccbar} reported 
the observation of $\etac$, $\chiczero$, and $\etacp$ 
in the mass distribution of the system recoiling 
against a reconstructed $\jpsi$ in $\ee$ annihilations. 
The production cross sections measured by Belle are about 
one order of magnitude higher than those predicted by 
non-relativistic QCD (NRQCD) 
calculations~\cite{ct:singleGamma-Chao,
                   ct:singleGamma-Braaten-Lee,ct:singleGamma2-Chao} 
for $\ee\to\gamma^*\to\jpsi\,\ccbar$ reactions, 
where $\ccbar$ is a charmonium state with even C-parity. 
There have been 
attempts~\cite{ct:Zhang-JpsiEtac, ct:Bondar-JpsiEtac, ct:Bodwin-twoPhoton, 
               ct:Bodwin-JpsiJpsi, ct:Brodsky-Glueball} 
to reconcile the large discrepancy between 
the observed cross section and predictions, and the validity of NRQCD 
approximations has been questioned~\cite{ct:Bondar-JpsiEtac, ct:Brambilla-QWG2004}. 
It has also been suggested that at least part of 
the double charmonium production 
might be due to two virtual-photon 
interactions~\cite{ct:Bodwin-twoPhoton}, 
$i.e.$, $\ee\to\gamma^{*}\gamma^{*}\to\jpsi\,\ccbar$, 
where odd C-parity states could be produced. 
Belle updated its observation and explored the origin of
the $\jpsi\,\ccbar$ events~\cite{ct:Belle-double_ccbar-update}.

In this paper we present a measurement of the cross sections 
for $\ee\to\jpsi\,\etac$, $\ee\to\jpsi\,\chiczero$, 
and $\ee\to\jpsi\,\etacp$, 
and set limits on the yields for other known charmonium states 
produced in association with a $\jpsi$. 
We calculate the mass ($M_\textrm{rec}$) of the system
recoiling against a fully reconstructed $\jpsi$ via:
\begin{equation}
  \mrec^2 = (\sqrt{s}-E^*_{\jpsi})^2 - p^{*2}_{\jpsi}\,, 
\label{eq:mrec}
\end{equation}
where $\sqrt{s}$ is the $\ee$ annihilation energy in the center-of-mass (CM) 
system, and $E^*_{\jpsi}$ and $p^*_{\jpsi}$ are the energy and momentum 
of the $\jpsi$ candidate in the CM system.


In this paper, we analyze 
112\invfb of data collected at the peak of the $\Y4S$ resonance and 
12\invfb at $\sqrt{s}=10.54\gev$, just below the $\Y4S$, 
with the \BaBar detector~\cite{ct:BaBar-detector} 
operating at the asymmetric energy PEP-II $\ee$ storage ring. 
The \BaBar detector includes a five-layer, 
double-sided silicon vertex tracker (SVT) 
and a 40-layer drift chamber (DCH) in a 1.5-T solenoidal magnetic field, 
which detects charged particles and measures their momenta 
and specific ionizations (\dedx). 
Photons and electrons are detected with a CsI(Tl)-crystal 
electromagnetic calorimeter (EMC). 
An internally reflecting ring-imaging Cherenkov (DIRC)
is used for particle identification. 
Penetrating muons are identified by an array of 
resistive-plate chambers (RPC) embedded in 
the steel of the flux return (IFR).


We select events with at least five well reconstructed charged tracks in 
the DCH, within the fiducial volume 0.41$< \theta<$2.54, where $\theta$ 
is the polar angle. 
Electron candidates have 
a pattern of specific ionization (\dedx) in the DCH, 
a Cherenkov cone angle, an EMC 
shower energy divided by momentum, and a number of EMC crystals that 
are consistent with an electron hypothesis.
A muon candidate is selected on the basis of energy deposited in 
the EMC, the number and distribution of hits in the IFR, 
and the match between the IFR hits and the extrapolation 
of the DCH track into the IFR. A more detailed explanation 
of particle identification is given elsewhere ~\cite{ct:BaBar-jsi}.

A pair of oppositely charged lepton candidates originating from a common
vertex is selected as a $\jsi$ candidate if its mass ($m(\ll)$) 
falls within [-50,30]\mevcc (for $\ee$) or [-30,30]\mevcc (for $\mm$), 
of the nominal $\jsi$ mass of 3.097\mevcc~\cite{ct:PDG2004}. 
In the calculation of $m(\ee)$, 
electron candidates are combined with nearby photon candidates 
in order to recover some of the energy lost through 
bremsstrahlung radiation. 
These mass intervals are referred to as the $\jpsi$ mass windows.
In order to improve the $p^*_{\jpsi}$ resolution, we perform a kinematic 
fit where the $\jsi$ candidate is constrained to have the nominal 
$\jpsi$ mass. 

There are two main background sources in this analysis: events with 
genuine $\jpsi$ mesons and combinatorial background. 
The region 60\mevcc$<|M(\ll)-M(\jpsi)|<$200\mevcc, 
defined as the $\jpsi$ mass sidebands, 
where $M(\jpsi)$ is the nominal $\jsi$ mass, 
is used to estimate the combinatorial background 
due to random tracks. 
This background is largely 
rejected by particle identification, and by a requirement on the lepton 
helicity angle in the $\jsi$ decay, $\abs{\cosHel}<0.9$, 
as shown in Fig.~\ref{fg:cut-vars}(a) and \ref{fg:cut-vars}(c). 

The largest backgrounds are due to real $\jpsi$ mesons from
QED processes such as $\jpsi$ or $\psip$ mesons 
produced via initial state radiation (ISR).
$\jpsi$ mesons from $B$ meson decay have $p^*<2\gevc$ 
and do not constitute 
a background for recoil masses below 6.6\gevcc.
Most QED backgrounds have low multiplicity, and 
may have electrons or photons escaping detection along the beam line.
These backgrounds are suppressed by the requirement of 
at least five charged tracks and 
the following requirement: for each event we calculate the 
energy deposited in the EMC plus the energy that can 
be attributed to an undetected electron or photon, 
\begin{equation}
  \Eqed= E_\textrm{EMC} + p_\textrm{miss}\,,
\end{equation}
where $E_\textrm{EMC}$ is the total energy deposited in the EMC, 
and $p_\textrm{miss}$ is the missing momentum in the lab frame in the event. 
We require $\emcTotLess<-1.0$\gev as shown in  
Fig.~\ref{fg:cut-vars}(b) and \ref{fg:cut-vars}(d), 
where $E_\textrm{beams}$ is the sum of the $\ee$ beam energies calculated 
in the lab frame. 
\bfg[htbp]
\infg[height=4.7cm,width=8.5cm]{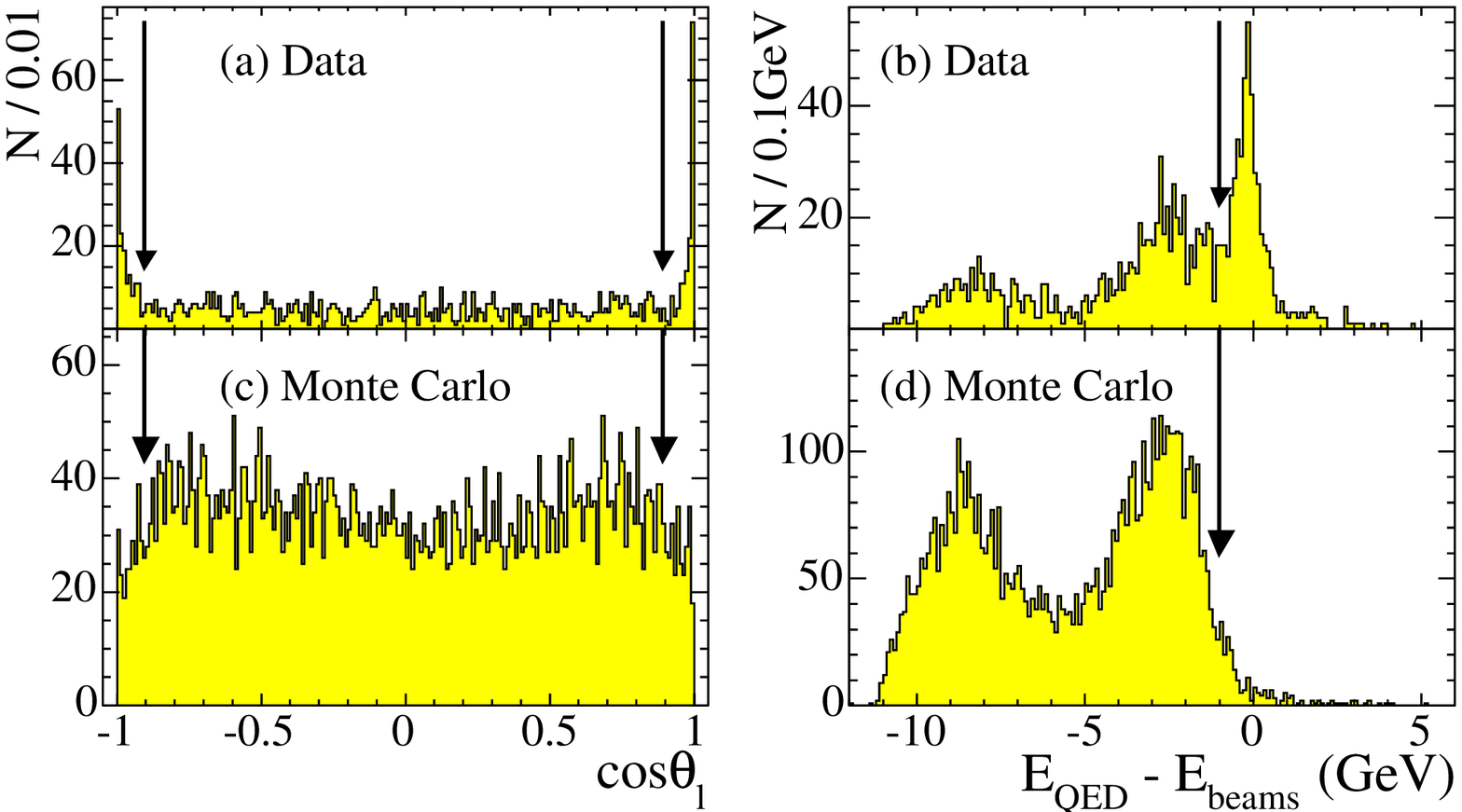}
\caption{Distributions of 
  (a) $\cosHel$ and (b) $\emcTotLess$ in the data, 
  (c) $\cosHel$ and (d) $\emcTotLess$ in the signal Monte Carlo.  
   The arrows point to where the selection criteria are applied.}
\label{fg:cut-vars}
\efg
We reject the $\jsi$ background from $\psip$ events by vetoing  
events if the invariant mass of the 
$\jpsi$ candidates combined with any pair of oppositely charged tracks 
with pion mass hypothesis 
is within 15\mevcc of the $\psip$ mass.

The recoil mass distribution for events in the $\jpsi$ mass window 
is shown as points with error bars in Fig.~\ref{fg:fit_mrec-jsi}. 
The ISR $\psip$ background is estimated using a Monte Carlo sample of 
ISR $\psip$ events. 
The $\psip$ feeddown background from continuum production is estimated 
using continuum $\psip$ events selected in the data.

\bfg[htbp]
  \centering
  \infg[height=4.7cm,width=8cm]{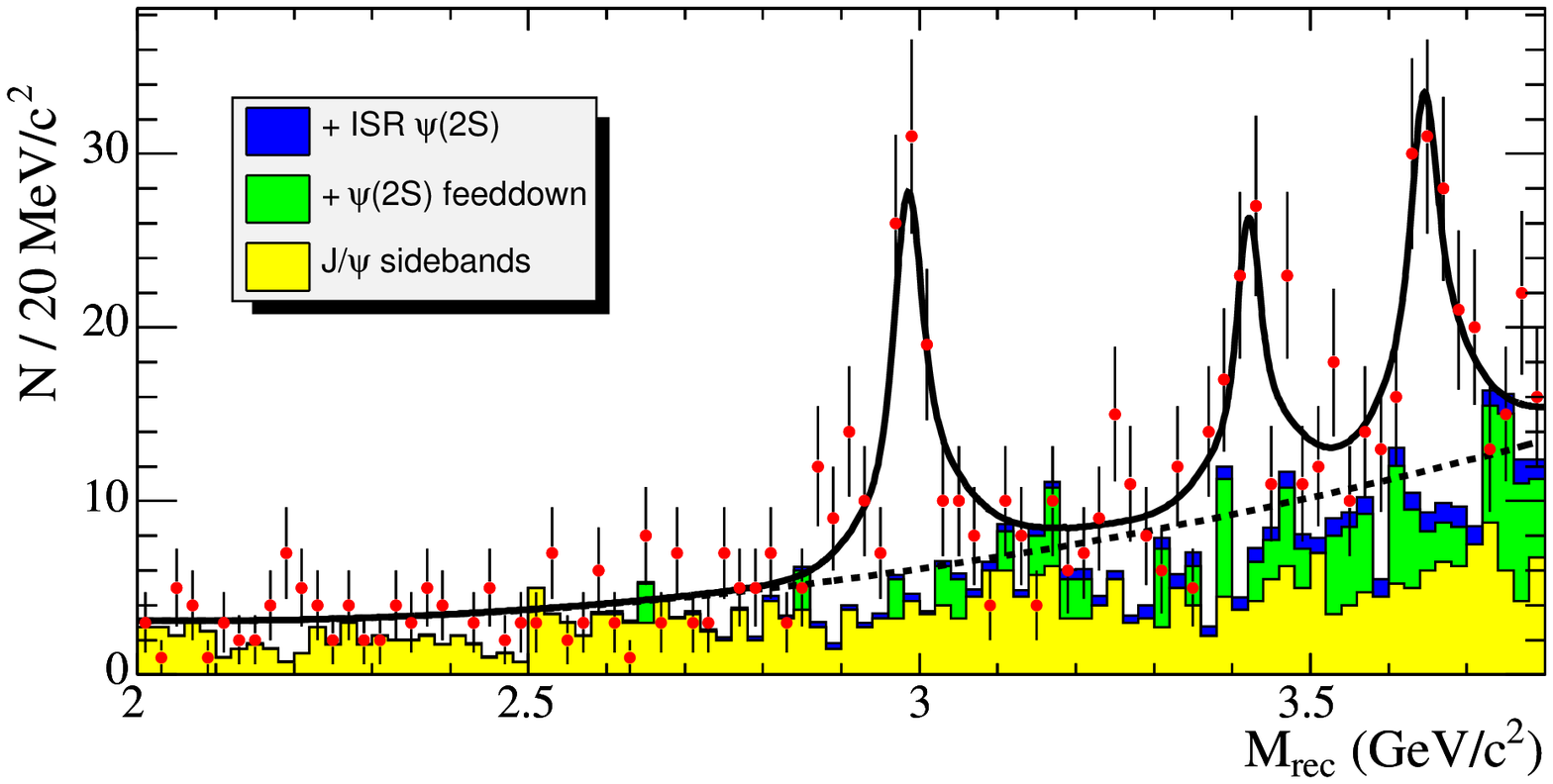}
  \label{fg:sum-std_fit_mrec-jsi}
  \caption{The fit to the recoil mass distribution is represented by 
     the solid curve. 
     The dashed curve is a second-order polynomial representing the background.
     The points with error bars refer to the events in the $\jsi$ mass window.
     The histograms represent different sources of backgrounds.}
\label{fg:fit_mrec-jsi}
\efg


The spectrum in Fig.~\ref{fg:fit_mrec-jsi} is fit 
to the sum of signal functions representing 
the $\etac$, $\chiczero$, and $\etacp$ lineshapes, plus 
a second-order polynomial background function. 
 The signal line shapes are obtained by convoluting the Breit-Wigner 
 line shape of each resonance with a fixed-width Gaussian representing the 
recoil mass resolution function. The widths of the Gaussians are determined 
from a Monte Carlo simulation of the momentum of the reconstructed $\jsi$; 
the $\jsi$ momentum resolution is different for the $\jee$ and $\jmm$ 
samples, but independent of the recoiling system.
This shape in turn is convolved with a long radiative tail that is 
calculated to $\order(\alpha^2)$~\cite{ct:generator-kk2f} 
for ISR photons that carry off an energy greater than 10 MeV.
The free parameters in the data fit are the coefficients for the background 
parameterization, the event yields for each resonance,  
the masses of the resonances, and the $\etacp$ total width. 
The total widths for the $\etac$ and the $\chiczero$ are fixed to
their world average values~\cite{ct:PDG2004} 
of 17.3\mevcc and 10.1\mevcc, respectively. 
The fit is performed simultaneously to the recoil mass spectra in the 
$\jee$ and $\jmm$ samples, and the total event yield for each 
resonance is given by the sum of the yields in each mode.

The fit result is given in Table~\ref{tb:fit-results} 
and is shown as the solid curve in Fig.~\ref{fg:fit_mrec-jsi}.
Other known charmonium states may also be produced in association with the
$\jpsi$ via two virtual-photon interactions. 
We therefore attempt to
include in our primary fit each one of the other known charmonium 
resonances in turn to determine their event yields, 
which are presented in Table~\ref{tb:fit-results}. We find no evidence 
for $\jpsi$, $\chi_{c1}$, $\chi_{c2}$, or $\psip$ in the mass spectrum 
of the system recoiling against a $\jpsi$.

\btbl
\centering
\caption{Result of the fits to the recoil-mass spectrum.  
  The errors are statistical only. 
  Where indicated, the value of the corresponding
  parameter is fixed to the current world average~\cite{ct:PDG2004}. 
  The primary fit is obtained including signals of $\etac$, $\chiczero$,
  and $\etacp$. The event yield for the other resonances 
  is determined by including each resonance in the primary fit.}
\label{tb:fit-results}
\btbu{p{1.9cm}ccc}
\hline \hline
  Recoil              & Number        & Mass              & Total Width         \\ 
  System              & of Events     & (\mevcc)          & (\mevcc) \\ 
 \hline 
 $\etac$              & $126\pm20$    & ~~~~$2984.8\pm4.0$    & fixed\\                          
 $\chiczero$          & $~81\pm20$    & ~~~~$3420.5\pm4.8$    & fixed\\ 
 $\etacp$             & $121\pm27$    & ~~~~$3645.0\pm5.5$  & $22\pm14$\\
 $\jpsi$              & $-26\pm13$    &  fixed            & fixed\\
 $\chi_{c1}$          & $ -5\pm16$    &  fixed            & fixed\\
 $\chi_{c2}$          & $-12\pm16$    &  fixed            & fixed\\
 $\psip$              & $~30\pm27$    &  fixed            & fixed \\
\hline \hline
\etbu
\etbl

The topological branching fraction 
is unknown for the $\etac$, $\chiczero$, and $\etacp$, so we report 
the product of the branching fraction 
for final states with more than two charged tracks 
($\BR_{>2}(\ccThreeChg)$) times  
the double charmonium production  cross section. 
In order to include the effect of ISR, the yields
reported in Table~\ref{tb:fit-results} are calculated with a 
 line shape based on a model of the $\sqrt{s}$ dependence of double charmonium
production model. 
To allow a direct comparison of experimental results, 
we follow the same method used by Belle~\cite{ct:Belle-double_ccbar-update} 
to remove this model dependence by determining cross section values 
that correspond to the non-tail fraction of the fit shape 
($f_{rad}=0.61$)~\cite{ct:generator-kk2f} 
where no ISR photon with an energy greater than 10\mev is radiated. 
We use
\begin{equation}
   \sigma(\ee\to \jpsi\,\ccbar)\,\BR_{>2} =   
   \frac {N_{\ccbar}f_{rad}} {\BR(\jpsi\to\ll)\,\LUMI\,\eff} \,,
\end{equation}
where $N_{\ccbar}$ is the event yield, 
      $\LUMI$ is the integrated luminosity, 
      $\BR(\jpsi\to\ll)$ is the $\jpsi$ branching fraction, 
      and $\eff$ is the detection efficiency. 
The value of $\eff$ is determined using a Monte Carlo simulation 
with the assumption that exclusive $\jpsi\eta_c(nS)$ production is $P$ wave 
and that exclusive $\jpsi\,\chiczero$ production is $S$ wave, 
as expected for a single virtual-photon process.
The efficiency is determined to be (28.8$\pm$0.7)\% for the $\etac$,
(31.5$\pm$0.7)\% for the $\chiczero$, and
(28.9$\pm$0.8)\% for the $\etacp$.

The systematic error is estimated taking into account contributions from 
the event selection and the fitting procedure, the particle identification 
efficiency, and the recoil-mass scale uncertainty.
The contributions from uncertainties in integrated luminosity 
and $\jpsi$ branching fraction are negligible. 
The contributions from individual sources 
(listed in Table~\ref{tb:incl-sys-error}) are 
added in quadrature, 
except for the systematic errors due to the mass-scale uncertainty, 
which are added linearly, 
to determine the total systematic errors. 

\btbl
\centering
\caption{Summary of systematic errors: variations of cross sections 
             and masses due to the selection and fitting procedure (Fit), 
             particle identification (PID) efficiency, 
             and recoil-mass scale uncertainty. 
             $\Delta\,M$ refers to the mass difference 
             between the $\etacp$ and $\etac$. }
\label{tb:incl-sys-error}
\btbu{cccc|cccc}
\hline \hline
   & \mcol{3}{Variations($\%$)}
             & \multicolumn{4}{c}{Variations(\mevcc)} \\
   & \mcol{3}{in Cross-section}
             & \multicolumn{4}{c}{in Mass} \\
Source             & $\etac$          & $\chiczero$       & $\etacp$ 
             & $\etac$          & $\chiczero$       & $\etacp$           & $\Delta\,M$\\
\hline 
Selection    & $^{+3.5}_{-8.3}$  & $^{+0.3}_{-9.2}$  & $^{+12.6}_{-15.6}$  
             & $^{+3.0}_{-0.2}$  & $^{+1.2}_{-0.7}$  & $^{+1.0}_{-5.1}$   & $-7.2$ \\[2mm]

Fit          & $^{+6.7}_{-8.1}$  & $^{+13.5}_{-14.2}$  & $^{+6.8}_{-8.3}$ 
             & $^{+0.1}_{-3.4}$  & $^{+9.9}_{-8.0}$   & $^{+3.3}_{-3.7}$ & $^{+7.1}_{-2.3}$ \\[2mm]

PID          &  $\pm3.5$  &  $\pm3.5$   &  $\pm3.5$
             &  -         &  -          &  -         &  - \\

Mass Scale   &  -         &  -          &  -
             & $\pm1.5$   &  $\pm1.5$   &  $\pm1.5$  &  0 \\
\hline 
Sum           & $^{+8}_{-12}$    &  $^{+14}_{-17}$     &  $^{+15}_{-18}$  
              & $^{+4.5}_{-5.0}$ &  $^{+11.5}_{-9.5}$  &  $^{+4.9}_{-7.8}$   & $^{+7.1}_{-7.6}$\\
\hline \hline
\etbu
\etbl

We obtain $\sigma(\ee\to \jpsi\,\ccbar)\,\BR(\ccThreeChg)$ to be  
    $17.6\pm2.8^{+1.5}_{-2.1}$~fb for $\jpsi\,\etac$, 
    $10.3\pm2.5^{+1.4}_{-1.8}$~fb for $\jpsi\,\chiczero$, 
and $16.4\pm3.7^{+2.4}_{-3.0}$~fb for $\jpsi\,\etacp$.
Throughout this paper, the first error is statistical and the second systematic.
Our values of the cross sections are consistent with 
Belle's measurements~\cite{ct:Belle-double_ccbar-update} 
for all three resonances. 
The cross sections measured by both experiments are 
much larger than those predicted by many NRQCD calculations. 

\btbl
\centering
\caption{Comparison of cross-sections ($\sigma\times\BR_{>2}$ in fb)
             with Belle's results~\cite{ct:Belle-double_ccbar-update}, 
             and with theoretical expectations 
             that do not include the $\BR_{>2}$ factor. 
             }
\label{tb:crossSection-comparison}
\btbu{cccc}
\hline \hline
 $\jpsi\,\ccbar$     & $\etac$     & $\chiczero$  & $\etacp$     \\ 
\hline
 \BaBar
                    & $17.6\pm2.8^{+1.5}_{-2.1}$  &  $10.3\pm2.5^{+1.4}_{-1.8}$ &  $16.4\pm3.7^{+2.4}_{-3.0}$  \\
  Belle~\cite{ct:Belle-double_ccbar-update}
                    & $25.6 \pm2.8 \pm3.4$  &  $ 6.4 \pm1.7 \pm1.0$ &  $16.5 \pm3.0 \pm2.4$   \\

NRQCD ~\cite{ct:singleGamma-Braaten-Lee}   & $2.31\pm1.09$         &  $2.28\pm1.03$        &  $0.96\pm0.45$       \\

NRQCD ~\cite{ct:singleGamma-Chao}          & $5.5$                 &    $6.9$              &  $3.7$  \\
\hline \hline
\etbu
\etbl

From the fit to the recoil mass spectrum we determine the 
$\eta_{c}(2S)$ mass to be $3645.0\pm5.5^{+4.9}_{-7.8}$\mevcc,  
and the total width to be $22\pm14$\mevcc.
The systematic errors are mainly due to the uncertainty 
on the $\jsi$ momentum measurement. 
We use ISR $\jsi$ and ISR $\psip$ data samples to determine the 
momentum shifts away from the expectations for ISR events. 
Assuming a constant momentum shift, we obtain the recoil mass uncertainty 
for $\jsi\,\ccbar$ processes due to the $\jsi$ momentum uncertainty. 
The mass difference ($\Delta M$) between the $\etacp$ and $\etac$ 
does not significantly depend on the absolute momentum scale 
and common systematic errors mostly cancel. 
We measure $\Delta M=660.2\pm6.8^{+7.1}_{-7.6}$\mevcc, 
which is in good agreement with 
the mass difference previously reported by 
this experiment~\cite{ct:BaBar-yy-etac2S} 
and by other experiments~\cite{ct:CLEO-yy-etac2S,ct:Belle-double_ccbar-update}.


In summary, we have measured the cross section for double 
charmonium production 
$\sigma(\ee\to \jpsi\,\ccbar)\,\BR(\ccThreeChg)$ 
for $\jpsi\,\etac$,  $\jpsi\,\chiczero$, 
and $\jpsi\,\etacp$. 
We confirm the unexpectedly large cross 
sections previously reported by the Belle experiment for these processes. 
No evidence is found for $\ee\to \jpsi\,\jpsi$, $\jpsi\,\chi_{c1}$, 
$\jpsi\,\chi_{c2}$, or $\jpsi\,\psip$. 
We also measure the mass difference between the $\etacp$ and the $\etac$.

We are grateful for the 
extraordinary contributions of our \pep2\ colleagues in
achieving the excellent luminosity and machine conditions
that have made this work possible.
The success of this project also relies critically on the 
expertise and dedication of the computing organizations that 
support \babar.
The collaborating institutions wish to thank 
SLAC for its support and the kind hospitality extended to them. 
This work is supported by the
US Department of Energy
and National Science Foundation, the
Natural Sciences and Engineering Research Council (Canada),
Institute of High Energy Physics (China), the
Commissariat \`a l'Energie Atomique and
Institut National de Physique Nucl\'eaire et de Physique des Particules
(France), the
Bundesministerium f\"ur Bildung und Forschung and
Deutsche Forschungsgemeinschaft
(Germany), the
Istituto Nazionale di Fisica Nucleare (Italy),
the Foundation for Fundamental Research on Matter (The Netherlands),
the Research Council of Norway, the
Ministry of Science and Technology of the Russian Federation, and the
Particle Physics and Astronomy Research Council (United Kingdom). 
Individuals have received support from 
CONACyT (Mexico),
the A. P. Sloan Foundation, 
the Research Corporation,
and the Alexander von Humboldt Foundation.

\end{document}